\RequirePackage{fix-cm}
\documentclass[smallextended]{svjour3}       
\smartqed  
\usepackage{graphicx}
\begin{document}

\title{$\Lambda(1405)$ observations in p+p and K$^-$-induced reactions}


\author{L. Fabbietti, K. Pischicchia, A. Scordo}


\institute{L.Fabbietti \at
               E12, Physik Department Technische Universität München, Germany\\
                Excellence Cluster "Origin and Structure of the Universe', Technische Universität München, Garching,Germany\\
                        \email{laura.fabbietti@ph.tum.de}            \\
                        K. Piscicchia \at		
                       INFN, Laboratori Nazionali di Frascati, Frascati (Rome), Italy \\
                       Centro Fermi - Museo Storico della Fisica e Centro Studi e Ricerche "Enrico Fermi", (Rome), Italy \\
                        \email{kristian.piscicchia@gmail.com}\\
                        A. Scordo \at	
                         INFN, Laboratori Nazionali di Frascati, Frascati (Roma), Italy	\\
                        \email{alessandro.scordo@lnf.infn.it}                   
          }

\date{Received: date / Accepted: date}

\maketitle

\begin{abstract}
The $\Lambda(1405)$ production in p+p collisions at 3.5 GeV and K$^-$-induced reactions is discussed.
The shift of the measured spectral function of the $\Lambda(1405)$ in p+p reactions does not match either theoretical
calculations for p+p reactions or experimental observation in previous  K$^-$-induced reactions. New experiments 
with stopped and in-flight $K^-$ are needed to
study this initial state more in detail. The state of the art of the analysis is discussed.

\end{abstract}

\section{Introduction}
Lying slightly below the $\mathrm{\bar{K}-N}$ threshold ($\approx30\,\mathrm{MeV/c^2}$), the broad $
\Lambda(1405)$ resonance is considered to be linked to the antikaon-nucleon interaction. Hence the understanding 
of this resonance is mandatory to address the issue of the interaction. From a theoretical point of view the $
\Lambda(1405)$ is treated within a coupled channel approach, based on chiral dynamics, in which the low-energy $
\mathrm{\bar{K}-N}$ interaction can be handled \cite{theoSumm}. 
In this Ansatz the $\Lambda(1405)$ appears naturally as a dynamically generated resonance, resulting from the 
superposition of two components:
 a quasi-bound $\mathrm{\bar{K}-N}$ state and a $\Sigma\pi$ resonance.\\
At present, the molecule-like character of the $\Lambda(1405)$ is commonly accepted. However, the contribution of 
the $\Sigma\pi$ channel to the formation process is still discussed controversially. 

In general, models can be constrained above the $\mathrm{\bar{K}-N}$ threshold by $\mathrm{K^-p}$ scattering 
data and by the measurements of the $\mathrm{\bar{K}p}$, $\mathrm{\bar{K}n}$ scattering lengths extracted from 
kaonic atoms \cite{sid:2011,weis:2011}. 
Below threshold, the only experimental observable related to the $\mathrm{\bar{K}-N}$ interaction is the $
\Lambda(1405)$ spectral shape extracted from the decays $\Lambda(1405)\rightarrow{\approx100 \%} (\pi
\Sigma)^0$. 
In this work the measurements of the $\Lambda(1405)$ spectral function extracted from p+p collisions at 3.5 GeV 
measured by HADES at GSI and the ongoing analysis of the data collected with the KLOE apparatus at DA$\phi$NE 
\cite{Cur14} for kaon-induced reactions are discussed.
p+p collisions represent a rather complicated initial condition for the detailed calculation of the $\Lambda(1405)$ 
production and moreover the data published by the HADES collaboration \cite{L1,L2,Aga3} refer to the decays $
\Lambda \rightarrow \Sigma^{\pm}$ where the contribution of the $\Sigma(1385)^0$ \cite{Epp12,Aga4} can not be 
filtered out.

The $\Lambda(1405)$ production in $\overline{K}N$ reactions is of particular interest due to the prediction, in chiral 
unitary models \cite{theoSumm}, of two poles emerging in the scattering 
amplitude (with $S=-1$ and $I=0$) in the region of the  $\Lambda(1405)$ mass. One pole is located at higher 
energy with a narrow width, and is mainly coupled to the $\overline{K}N$ 
channel, while a second lower mass and broader pole is dominantly coupled to the $\Sigma \pi$ channel 
\cite{weise2012}. Both contribute to the final experimental invariant mass distribution \cite{oller,nacher}.
Moreover, the $\Sigma^0 \pi^0$ channel, which is free from the $I=1$ contribution and from the isospin interference 
term, represents the golden decay channel for such state.
One therefore expects two very different spectral function for the measured $\Lambda(1405)$ when employing 
proton or kaon beams.
In this work we revise the results published by the HADES collaboration where the $\Lambda(1405)$ was measured 
in p+p collisions at 3.5 GeV and we discuss the yet unpublished data measured with KLOE and stopped/in-flight 
antikaons impinging on carbon and helium targets.
\begin{figure}[hbp]
	\centering
		\includegraphics[width=0.65\textwidth]{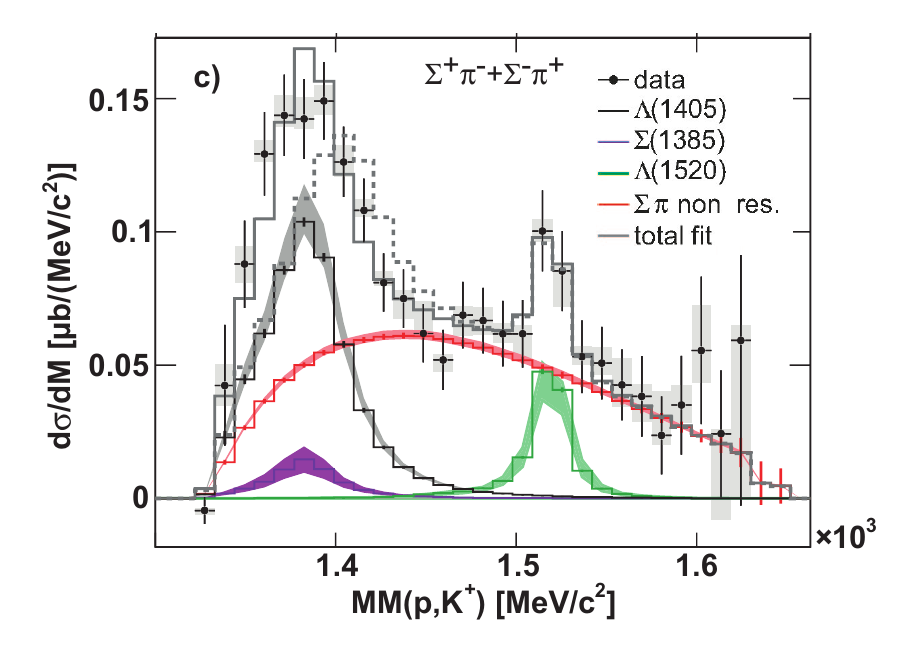}
	\caption{(Color online) Missing mass $\mathrm{MM(p,K^+)}$ distributions for events attributed to the $\Sigma^+\pi^-$ and $\Sigma^-\pi^+$ decay channels \cite{L1}. 
	See text for details.}
	\label{L1405HADES}
\end{figure}  

\section{Results from p+p collisions}
The HADES experiment at GSI is a multipurpose spectrometer with a large geometrical acceptance around mid-
rapidity for nucleon-nucleon and nucleus-nucleus collisions in a fixed target configuration at intermediate energies (E
$_{\mathrm{KIN}}=\,1-3.5$ GeV/nucleon).
Figure \ref{L1405HADES} shows the missing mass $\mathrm{MM(p,K^+)}$ distributions for events containing 
additionally a $\Sigma^+\pi^-$ or a $\Sigma^-\pi^+$ pair \cite{L1} extracted 
from p+p collisions at 3.5 GeV beam kinetic energy and measured with the HADES spectrometer.
The intermediate $\Sigma^+$ and $\Sigma^-$ hyperons have been reconstructed via the missing mass to the 
proton, the $\mathrm{K^+}$ and either the $\pi^-$ or the $\pi^+$  (see Fig.~4 in \cite{Valencia2010}). The details 
about the track reconstruction and particle identification are provided in \cite{Valencia2010,L1,Aga4}. The black 
symbols show the efficiency and acceptance corrected experimental data together with statistical and systematic 
errors (gray boxes), the black, violet and green continuous lines refer to the full scale simulation of the 
$\Lambda(1405),\,\Sigma(1385)^0$ and $\Lambda(1520)$ production in the
same final state, respectively. The $\Lambda(1405)$ has been simulated assuming a Breit-Wigner mass distribution 
peaked at $1385$ MeV/c$^2$ and with a 
width of $50$ MeV/c$^2$, the nominal PDG mass values have been used for the $\Sigma(1385)^0$ and 
$\Lambda(1520)$. The red line shows the simulation of the non-resonant 
contribution of the same final state. The grey line represents the sum of these simulations and reproduces 
the experimental data very well. When simulating 
the $\Lambda(1405)$ with a Breit- Wigner distribution peaking at $1405$ MeV/c$^2$ and leaving all the other 
components unvaried, the resulting sum of the simulated channels is represented by the grey dashed line. 
One can see that this assumption is not compatible with the experimental data.
The limited statistics of these data doesn't allow to distinguish between a Breit-Wigner and a Flatte' distribution for 
the $\Lambda(1405)$, despite of the fact that the latter function is more suited for the description of a molecular 
state. 

\begin{figure}[hbp]
	\centering
		\includegraphics[width=0.65\textwidth]{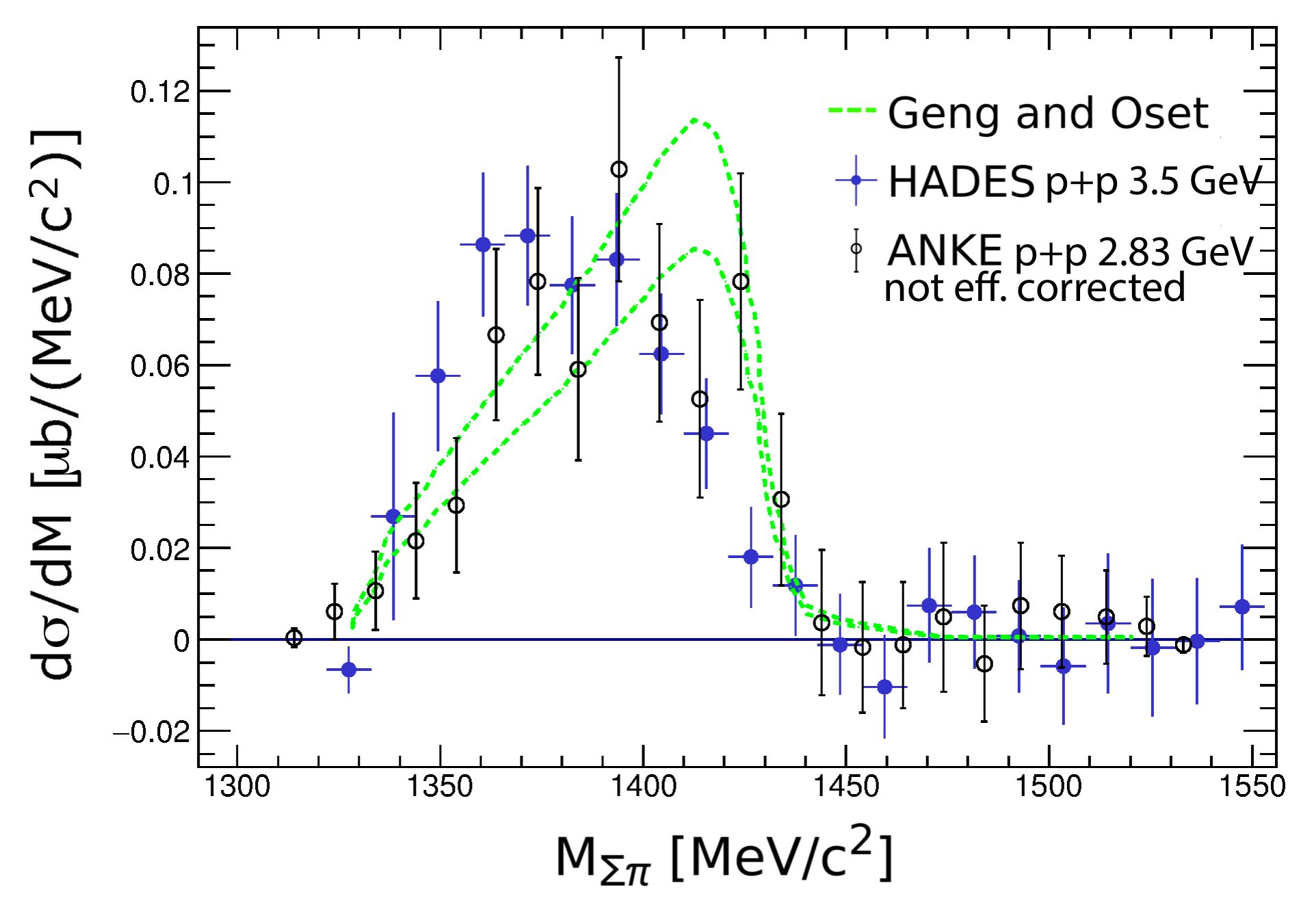}
	\caption{(Color online) Missing mass $\mathrm{MM(p,K^+)}$ distributions for events attributed to the HADES \cite{L1} and ANKE \cite{L1405Anke}
	data together with the $\Lambda(1405)$ spectral shape calculated in \cite{Gen08}. See text for details.}
	\label{L1405Comp}
\end{figure}

The differences expected in the spectral shapes of the $\Lambda(1405)$ reconstructed from the three decay 
channels $\Sigma^+\pi^-,\,\Sigma^-\pi^+,\,\Sigma^0\pi^0$ 
are not visible when comparing the two charged states measured by HADES (Fig. 1 in \cite{L1}). Also we know that 
the interference terms between the I=0 and I=1
states cancel out in the sum of the two charged amplitudes and that the remaining squared I=0 amplitude is 
dominant with respect to the squared I=1 amplitude.
This way, the sum of the $\Sigma^+\pi^-$ and $\Sigma^-\pi^+$ distributions should be equivalent to the $
\Sigma^0\pi^0$ distribution.
For this reason the sum of the two efficiency corrected charged decays of the $\Lambda(1405)$ measured by 
HADES  can be compared to the results published by the ANKE collaboration \cite{L1405Anke} and the 
corresponding theoretical prediction \cite{Gen08} where the neutral decay of the $\Lambda(1405)$ has been 
considered ($\Lambda(1405)\rightarrow \Sigma^0\pi^0$). 
The neutral decays offers the big advantage that it doesn't contain any contribution from the $\Sigma^0(1385)$ 
overlapping with the $\Lambda(1405)$ spectral shape, since the $\Sigma(1385)^0$
cannot decay into $\Sigma^0\pi^0$ pairs due to isospin conservation rules.
Figure \ref{L1405Comp} shows the comparison of the HADES data with the ANKE measurement and a theoretical calculation.
In order to compare the HADES data to the ANKE data and to the calculation by Geng and Oset the simulated 
curves corresponding to the sources other than the $\Lambda(1405)$ have been subtracted from the experimental 
missing mass $\mathrm{MM(p,K^+)}$ distribution shown in Fig.\ref{L1405HADES}.
This way, the pure $\Lambda(1405)$ spectral shape can be compared, under the assumption that no interferences 
occur among the different intermediate states.
The subtracted HADES spectrum is shown in Fig. \ref{L1405Comp} (full blue circles) together with the ANKE data 
(empty black circles) and the calculation (dashed green lines). 
Only statistical errors are displayed and the two dashed curves correspond to two different values of the total 
production cross-section for the $\Lambda(1405)$ in p+p collisions with a beam momentum of $3.65$ GeV/c 
($E_{KIN}=\,2.83$ GeV) of $4$ and $5.4\,\mu$b respectively. Indeed a total cross-section of $4.5\mu b$ for the
$\Lambda(1405)$ has been estimated from the ANKE analysis \cite{L1405Anke}.

The ANKE data are not efficiency-corrected but they have been compared to absolutely normalised theoretical 
calculation in \cite{Gen08}, where conclusions on the $\Lambda(1405)$ line shape have also been drawn. 
In Fig.~\ref{L1405Comp} the ANKE data are scaled such that the integral of the $\Sigma \pi$ invariant mass 
spectrum between 1320 and 1440 MeV/c$^2$ is the same as the integral obtained for the HADES distribution.
The theoretical calculations are also scaled by the same factor.
The different binning of the HADES and ANKE histograms is accounted for in the errors shown for the HADES 
distribution.
It is clear that here only a qualitative comparison of the two spectral shapes is possible, 
since the spectra published by ANKE are not corrected for efficiency and the beam
energies for the two experiments differ. It has also to be pointed out that the ANKE 
spectrum does not show any sign of the expected $\Lambda(1520)$, while the cross-section extracted for this state 
from the HADES data is about half of the $\Lambda(1405)$ production cross section \cite{L1}.
One can see that the two experimental distributions are in reasonable agreement even if this comparison is only 
qualitative. One can also see that the Geng-Oset calculations fail to reproduce the HADES data 
but also are compatible with the ANKE distributions only because of the large statistical error.
A direct comparison would be only possible if a dedicated calculation for the charge decays of the $\Lambda(1405)$ 
produced in p+p collisions at $3.5$ GeV kinetic energy would be available. 
Another crucial point is the presence of the non-resonant background, which is measured experimentally and might 
even interfere with the resonant contributions but is not considered yet in the calculation. Theoretical calculations as 
reported in \cite{Gen08} includes also the excitation of an intermediate  N$^*(1710)$ but actually also other 
resonances could be excited and the non-resonant contribution should also be taken into account for the case of p+p collisions at $3.5$ GeV.
The comparison of the available data from p+p collisions suggest a shift of the $\Lambda(1405)$ pole mass towards 
lower value than the assumed nominal mass at $1405$ MeV/c$^2$ but more detailed calculations should be carried 
out and compared to the data.

\subsection{Kaon-induced reactions with AMADEUS}
The AMADEUS experiment \cite{AMADE,AMADEUS} has the aim to perform studies of the low energy hadronic 
interactions of negatively charged kaons with nucleons and nuclei, which are fundamental to solve longstanding 
open questions in the non-perturbative QCD strangeness sector. The experiment is
 located at the DA$\Phi$NE collider that provides a unique source of monochromatic low-momentum kaons 
 stemming from the decay of the $\phi$ meson produced almost at rest within the $e^+-e^-$ collider. The antikaons 
 from the $\phi$ decay (BR$\approx 50\%$) with a momentum of about $140\,$MeV/c 
  can interact with the materials within the KLOE Drift Chamber (DC), that is used as an active target.
  This way, the DC provides an excellent acceptance and resolution data for $K^-$ capture  on H, ${}^4$He, ${}^{9}$Be and ${}^{12}$C, both at-rest and in-flight. 
  The presence of a electromagnetic calorimeter surrounding the DC allows also for photon identification.
  Among the many physics topics that can be addressed within the AMADEUS program, the study of the $\Lambda(1405)$ spectral shape produced in 
  in-flight and at-rest antikaon reactions on different target types is of crucial importance.
  In particular, the reconstruction of the decay $\Lambda(1405)\rightarrow \Sigma^0\pi^0$ can be addressed thanks to the particle identification
  capability of KLOE. Data collected in 2005 have been analysed to study the spectral shape of the $\Lambda(1405)$ reconstructed in at-rest and in-flight 
  reactions for K$^-$ impinging on carbon, helium and hydrogen targets.

\subsection{The $\Sigma^0 \pi^0$ identification}

The selection of $\Sigma^0 \pi^0$ events proceeds, after the $\Lambda(1116)$ identification, with the search for three additional photon clusters in time coincidence with each other
and stemming from the secondary decay vertex position reconstructed for  the  $\Lambda(1116)$ candidate ($\textbf{r}_\Lambda$). 
To this end, a minimisation of the variable $\chi_t^2=(t_i-t_j)^2/\sigma_t^2$, where $t_i$ represents the measured and corrected time of the  i-\emph{th} cluster,
is performed to select three neutral clusters in the calorimeter ($E_{cl}>20$ MeV).  According to dedicated MC simulations a cut was optimised 
on this variable $\chi_t^2 \le 20$.

\begin{figure}[tbp]
	\centering
		\includegraphics[width=0.93\textwidth]{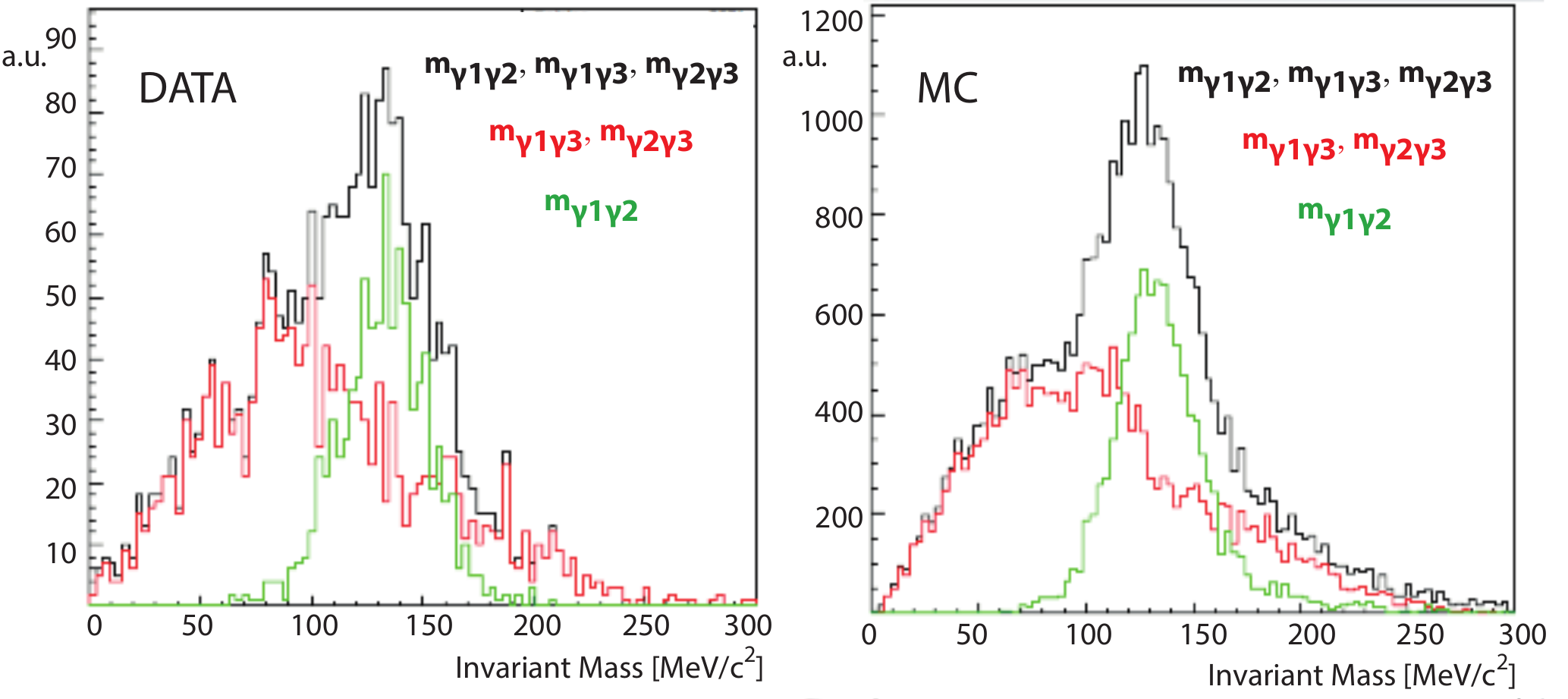}
	\caption{(Color online) $m_{\gamma \gamma }$ invariant mass distribution from experimental and simulated data for events with a $\Sigma^0$ candidate. See text for details.}
	\label{pi0}
\end{figure}

\begin{figure}[tbp]
	\centering
		\includegraphics[width=0.45\textwidth]{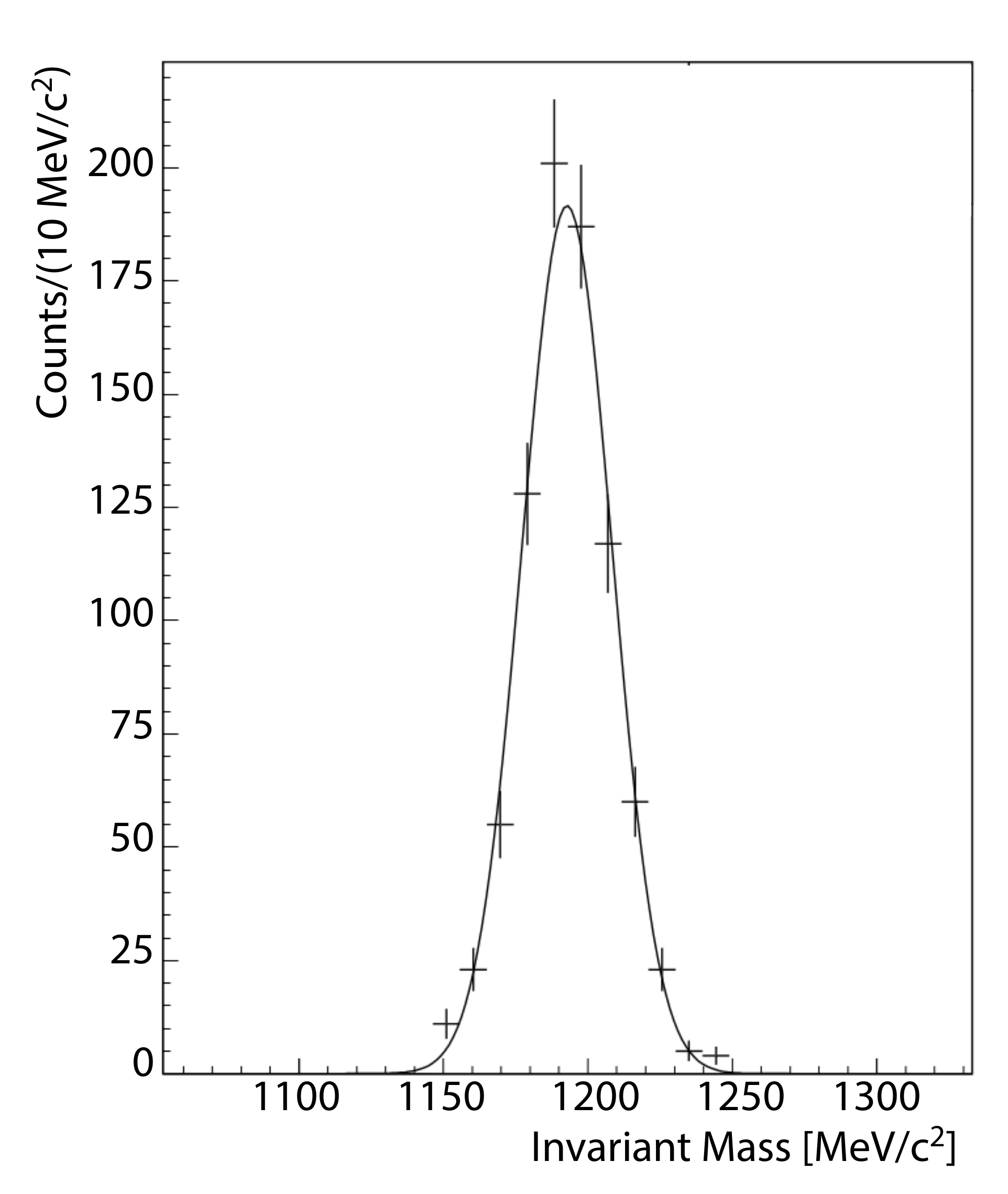}
	\caption{(Color online) $m_{\Lambda\gamma 3}$ invariant mass distribution, together with a Gaussian fit.}
	\label{plot}
\end{figure}

Once the three  photon cluster candidates are chosen, their assignment to the $\pi^0$ and $\Sigma^0$ decays is based on a second minimisation.  
Here, the variable $\chi_{\pi\Sigma} ^2$ is used, that involves both the $\pi^0$ and $\Sigma^0$ masses. 
This variable is calculated for each possible combination and the minimising triplet is selected. This selection was
optimised using MC simulations and the final selection corresponds to $\chi^2_{\pi\Sigma} \le 45$. 
According to true MC information the algorithm has an efficiency of $(98\pm1) \%$ in recognising photon clusters and an efficiency of $(78\pm1)\%$ in distinguishing the correct triplet.
A check is performed on the clusters energy and distance to avoid the selection of split clusters (single clusters in the calorimeter erroneously recognised as two clusters) for $\pi^0$s.
Figure \ref{pi0} shows the experimental and simulated invariant mass distribution for the photon pairs built from all the combination for the select minimising  photon triplet. The green distributions in both panels
show the invariant mass for the pair associated to the $\pi^0$ and the red curves correspond to the invariant mass of the 
other pairs. The black histogram represents the sum of all combinations within the minimising triplet.
 The $\gamma_1$ and $\gamma_2$ indexes indicate the two photons stemming from the $\pi^0$ decay, $\gamma_3$ refers
to the photon associated to the $\Sigma^0$ decay.
One can clearly see how the method allows for the reconstruction of the $\pi^0$ mass and also that the obtained experimental resolution is fairly well reproduced by the simulations.
Figure \ref{plot} shows the invariant mass $m_{\Lambda\gamma 3}$ (for absorptions in the DC volume) together with a Gaussian fit 
that delivers a mean value and a width ($\sigma$) of $1193$ and $15.65$ MeV/c$^2$ respectively. 
One of the major strengths of this analysis is the possibility of distinguish to a certain extend at-rest from in-flight reactions
via the measurement of the momenta of the pion and $\Sigma$ in the final state. The purity of this selection strongly
depends on the reacting target, so that in some cases overlaps are presents and hence contemporary fits on different
kinematic variables have to be employed.
\begin{figure}[tbp]
	\centering
		\includegraphics[width=0.75\textwidth]{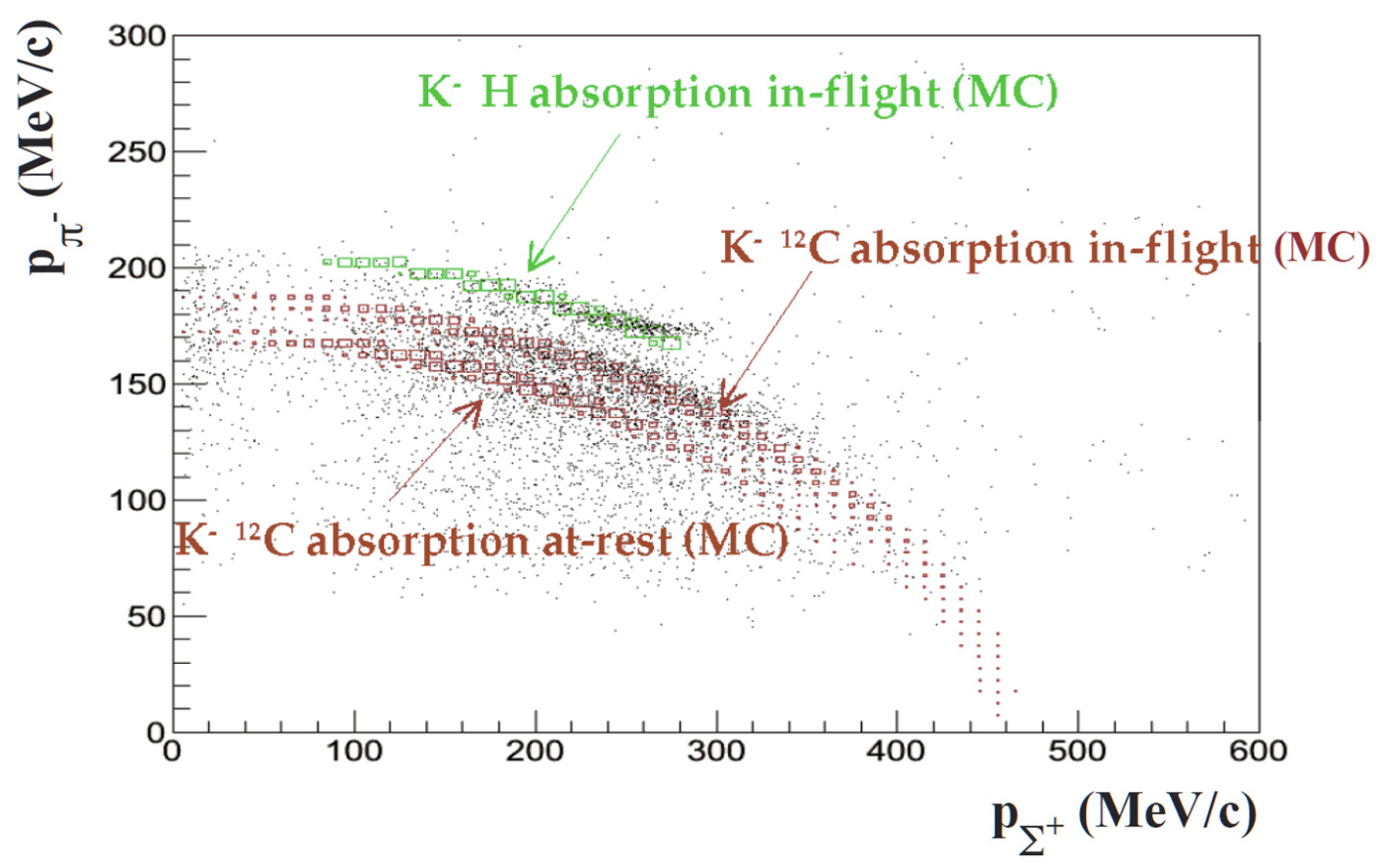}
	\caption{(Color online) $\pi^-$ momentum versus $\Sigma^+$ momentum. The black square represent the experimental distribution, the green
	symbol correspond to simulated $K^-+p$ at-rest reactions and the red symbols represent the simulated at-rest and in-flight reactions
	on carbon respectively.}
		\label{IfAr}
\end{figure}  
Figure \ref{IfAr} shows the pion versus $\Sigma$ momentum distribution for the selection of the two charged decay channels of the 
$\Lambda(1405)$. The experimental data are compared to full-scale simulation of at-rest and in-flight reaction on carbon (red symbols) and
at-rest reactions on hydrogen (green symbols). One can see that the different bands obtained from the simulations are well distinguishable
also in the experimental data. This example shows how the selection of the in-flight and at-rest reactions can be carried out.
Preliminary results which are still under final evaluation show that the peculiarity of the in-flight reactions is that
 higher $\Lambda(1405)$ are accessible with respect to the at-rest reactions.
 Final results are to be expected soon.
Since the measured $\Sigma^0\pi^0$ invariant mass distributions contain not only both the in-flight and at-rest reactions leading
to the $\Lambda(1405)$ formation on carbon and helium targets but also the non-resonant $\Sigma^0\pi^0$ production, a 
quantitative understanding of the the underlaying processes becomes mandatory. 
This require appropriate calculations of antikaon reactions with carbon and helium
nuclei including the correct treatment of spectator fragments and nucleons in order to describe the experimental \
momentum distribution correctly. This kind of calculation and comparison to the experimental data via full-scale simulations
are currently ongoing.  
\section{Summary}
We have discussed the $\Lambda(1405)$ results published by HADES comparing them directly to the ANKE measurement
and the calculation by Geng and Oset, that are to this end the only available for the p+p initial conditions. 
The HADES data are rather compatible with the ANKE measurement, whereas a quantitative comparison is at this stage not
possible since the ANKE data are not corrected for the geometrical acceptance and reconstruction efficiency.
The presence of the non-resonant background in the experimental data calls for more extended calculation that could include all
the contributing channels to the measured final states. \\
The feasibility of the measurement of the decay $\Lambda(1405)\rightarrow \Sigma^0\pi^0$ in reactions with monochromatic 
and low energy K$^-$ beam on helium and carbon target has been shown. Together with the excellent particle identification the
large acceptance of the KLOE spectrometer, employed in the phase 0 of the AMADEUS program, allows also to distinguish
between in-flight and at-rest processes and to hence study the $\Lambda(1405)$ spectral shape in different kinematical conditions. A comparison of the obtained line
shapes in the two reactions could help in better understanding the dynamic of the formation of the $\Lambda(1405)$ resonance.



\end{document}